\documentstyle[12pt,epsf]{article}
\newcommand{\be}{\begin{equation}}
\newcommand{\ee}{\end{equation}}
\newcommand{\ben}{\begin{eqnarray}\displaystyle}
\newcommand{\een}{\end{eqnarray}}

\newcommand{\p}{\partial}

\def\ben{\begin{equation}}
\def\een{\end{equation}}
\def\bena{\begin{eqnarray}}
\def\eena{\end{eqnarray}}

\input amssym.def
\input amssym.tex
\newcommand{\bk}{\hfill\break}

\def\fh{{\cal H}^+}
\def\ph{{\cal H}^-}
\def\finf{{\cal I}^+}
\def\pinf{{\cal I}^-}
\def\f{f_{\omega}}

\def\p{p_{\omega}}
\def\q{q_{\omega}}
\def\j{j_{\omega}}

\def\a{a_\omega}
\def\ad{a^{\dagger}_{\omega} }
\def\b{b_{\omega}}
\def\bd{b^{\dagger}_{\omega} }
\def\c{c_{\omega}}
\def\cd{c^{\dagger}_{\omega} }
\def\d{d_{\omega}}

\def\alp{\alpha _{\omega \omega '}}
\def\bet{\beta_{\omega\omega '}}
\def\vac{|0>_{in} }
\def\ub{\bar u}
\def\vb{\bar v}
\def\ubh{U_{bh}}

\def\uds{U_{ds}}
\def\vds{V_{ds}}
\def\kbh{\kappa_{bh}}
\def\kds{\kappa_{ds}}

\def\pw{p_\omega}
\def\bwwp{\beta_{\omega\omega^\prime}}
\def\awwp{\alpha_{\omega\omega^\prime}}
\def\bwwpbh{\bwwp^{bh}}
\def\bwwpds{\beta_{\omega\omega^\prime}^{ds}}
\def\nw{N_\omega}

\begin{document}
\begin{titlepage}
\vfill
%\begin{flushright} hep-th/0010055\\ \end{flushright}

%\centerline{\Large \bf {$\frak{D}\frak{R}\frak{A}\frak{F} \frak {T}$}}
%\centerline { \bf \today}

\vfill

\begin{center}
\baselineskip=16pt
{\Large\bf An Introduction to Black Hole Evaporation}
\vskip 0.3cm
{\large {\sl }}
\vskip 10.mm
{\bf Jennie Traschen } \\
%\\[2mm]
\vskip 1cm
%\vfill
{\small
  Department of Physics\\
  University of Massachusetts\\
  Amherst, MA 01003-4525\\
	 traschen@physics.umass.edu
}
\end{center}
\vfill
\par
\begin{center}
{\bf ABSTRACT}
\end{center}
\begin{quote}

Classical black holes are defined by the property that things can go in,
but don't
come out. However, Stephen Hawking calculated that black holes actually radiate
quantum mechanical particles. The two important ingredients that result in
back hole evaporation are (1) the spacetime geometry, in particular the
black hole horizon, and
(2) the fact that the notion of a ``particle" is not an invariant concept
in quantum field theory.
These notes contain a
step-by-step presentation of Hawking's calculation. We review portions of
quantum field theory in curved spacetime and basic results about static black
hole geometries, so that the discussion is self-contained.
Calculations are presented for quantum particle production for
an accelerated observer in flat spacetime, a black hole which forms from
gravitational collapse, an
eternal Schwarzschild black hole,  and charged black holes in asymptotically
deSitter spacetimes.  The presentation highlights the similarities in all
these calculations.
Hawking radiation from black holes also points to a profound connection
between black hole dynamics and classical thermodynamics.  A theory of
quantum gravity must predicting and
explain black hole thermodynamics.  We briefly discuss these issues
and point out a connection between black hole evaportaion and the positive mass
theorems in general relativity.

\vfill
%\hrule width 5.cm
%\vskip 2.mm

\end{quote}
\end{titlepage}
%%%%%%%%%%%%%%%%%%%%%%%%%%%%%%%%%%%%%%%%

\setcounter{equation}{0}

\noindent{\large\bf Table of Contents}
\vskip 0.05in

1. {\it Introduction}

2. {\it Quantum Fields in Curved Spacetimes}

3. {\it Accelerating Observers in Flat Spacetime}

4. {\it Black Holes}

5. {\it Particle Emission from Black Holes}

6. {\it Extended Schwarzchild and Reissner-Nordstrom-deSitter Spacetimes}

7. {\it Black Hole Evaporation and Positive Mass Theorems}

\section{Introduction}

Stephen Hawking published his paper ``Particle Creation by Black Holes"
\cite{hawking} in 1975.  In this article, Hawking demonstrated that classical black
holes radiate a thermal flux of quantum particles, and hence can be
expected to evaporate away.
This result was contrary to everything that was known about black holes and
classical
matter, and was quite startling to the physics community. However, the
effect has now been
computed in a number of ways and is considered an important clue in the
search for a theory
of quantum gravity.  Any theory of quantum gravity that is proposed must
predict black hole
evaporation.  The aim of these notes is to (1) develop enough of the
formalism of
semi-classical gravity to be able to understand the preceeding sentences,
excepting the term
``quantum gravity" itself, and (2) give a  step-by-step presentation of
Hawking's
calculation. We will also present a number of  related results on particle
production for an
accelerating observer in flat spacetime, and  for charged black holes in
asymptotically
deSitter spacetimes. Finally, we will discuss an interesting relationship
between classical positive
mass theorems in general relativity and endpoints of the quantum mechanical
process of Hawking evaporation.

For the record, Einstein's equation is given by
\be\label{einstein}G_{ab}\equiv R_{ab} -{1\over 2} g_{ab} R = 8\pi G_N
T_{ab} .\ee
Here $G_{ab}$ is the Einstein tensor, $R_{ab}$ the Ricci tensor, $R= R^a
_{\ a}$ is the scalar
curvature, $T_{ab}$ is the stress-energy tensor and $G_N$ is Newton's
gravitational constant. The
other constants of nature that come into the calculations are the speed of
light $c$ and
Planck's constant $\hbar$. In most of the paper, we will work in units with
$G=c=\hbar
=1$.

A black hole is a region in an asymptotically flat spacetime which is not
contained in the
past of future null infinity $\finf$.  The horizon is the
boundary between the black hole and the outside, asymptotically flat region.
In section (6) we will study a black hole in a spacetime which is not
asymptotically flat using an obvious generalization of the definition.
The horizon is a null surface. Physically, it is the outer boundary of the
black hole on which
null rays can just skim along, neither being captured by the black hole,
nor propagating
to null infinity.

Classical black hole mechanics can be summarized in following three basic
theorems, where the
necessary symbols are defined in section (4) below.\bk\vskip 0.05in\noindent
$0)$ The zeroth law states that the surface gravity $\kappa$ of a black hole is
constant on the horizon. \bk
$1)$ The first law states that variations in
the mass $M$, area $A$, angular momentum $L$, and charge $Q$ of a black hole
obey \cite{bch,carter}
\be\label{firstlaw}\delta M ={\kappa \over 8\pi } \delta A +\Omega \delta L
-\nu \delta Q,\ee
where $\Omega$ is the angular velocity of the horizon and $\nu$
is the difference in the electrostatic potential between infinity and
the horizon.\bk
$2)$ The second law is
the area theorem \cite{area} proved by Hawking in 1971. The area of a black
hole horizon is nondecreasing in time,
\be\label{areath}\delta A \ge 0\ee
This result assumes that the spacetime is globally
hyperbolic and that the energy condition $R_{ab}k^a k^b \ge 0 $ holds for all
null vectors $k^a$.\bk

These theorems bear a striking resemblance to the correspondingly numbered
laws of classical thermodynamics.  The zeroth law of thermodynamics says
that the temperature
$T$ is constant throughout a system in thermal equilibrium. The first law
states that in
small variations between equilibrium configurations of a system, the
changes in the energy $M$ and entropy $S$ of the system obey equation
\ref{firstlaw}, if
${\kappa \over 8\pi} \delta A$ is replaced by $T\delta S$, and the further
terms on the
right hand side are interpreted as work terms.  The second law of thermodynamics
states that, for a closed system, entropy always increases in any process,
$\delta S \ge 0$.

We see that the theorems describing black hole interactions, which are
results from
differential geometry, are formally identical to the laws of classical
thermodynamics, if one
identifies the black hole surface gravity $\kappa$ with a multiple of $T$ and the area of
the horizon $A$ with a multiple of the entropy $S$.   It is tempting to
wonder whether this
identification is more than formal.  Such a conjecture seems to require a
drastic
shift in the meaning of the geometrical properties of a black hole.
Temperature is a
measure of the mean energy of a  system with a large number, {\it e.g.}
order $10^{23}$, of
degrees of freedom. Entropy measures the number of microscopic ways these
degrees of
freedom can be arranged to give a fixed macroscopic configuration, {\it
e.g} fixed $M$, $L$
and $Q$. It is not at all obvious that the surface gravity and area of a
black hole should
have anything to do with a statistical system with a large number of degrees of
freedom.  Even more glaring, is the problem of radiation.  A hot lump of
coal radiates. And
the definition of a black hole is that it does not radiate; things go in,
but don't come out.

Nonetheless, in 1973 Bekenstein \cite{beck} suggested that a physical
identification does hold
between the  laws of thermodynamics and the laws of black hole mechanics.
Then in 1975,
Hawking published his calculation that black holes do indeed radiate, if
one takes into
account the quantum mechanical nature of matter fields in the spacetime.

\section{Quantum Fields in Curved Spacetimes}

\vskip 0.1in\noindent
{\it The Basic Idea of Particle Production}
\vskip 0.05in

The basic idea of semiclassical gravity is that, for energies below the
Planck scale,
it is a good approximation to treat matter fields quantum mechanically, but keep
gravity classical.  Hence, one considers quantum field theory in a fixed
curved background.
We will focus on free scalar field that classically satisfies the wave equation
\be\label{wave}g^{ab}\nabla _a \nabla _b \phi =0\ee
The scalar field $\phi$ is a quantum operator. This means that
(1) $\phi$ must obey the canonical equal time commutation relations
$[\phi (t, x^i ), \phi (t, y^i )] =\delta ^3 (x^i -y^i )$,
and (2) we must define a Hilbert space of states on which these operators act.
Physical observables are then computed by taking expectation values of the
corresponding operators in a given
state, or more generally matrix elements between states.

The key idea behind quantum particle production in curved spacetime is
that the definition of a particle is observer dependent.  It depends
on the choice of reference frame.  For example, an observer Al has a natural
time coordinate defined by proper time $T$ along Al's world line.
As we will discuss in more detail below, Al defines
particles as positive frequency oscillations of the scalar field with
repect to this time
$T$.  A second observer, Emily, will define particles as positive frequency
oscillations with
respect to her own proper time $t$. In general, the number of $T$-particles
that Al measures
will be different than the number of $t$-particles that Emily measures.
This effect occurs even in flat spacetime \cite{fulling,unruh}.
Since quantum field theory in flat spacetime is globally Lorentz invariant,
if Al
and Emily's frames differ only by a Lorentz transformation, then they
{\it will} agree about particle content.  However, if they have a
relative acceleration, then they will measure different particle numbers.
In the next section, we will study the case when Al uses
global inertial coordinates, while Emily undergoes constant acceleration.
We will see that in this case, when Al measures spacetime to be empty of
his $T$-particles,
Emily will measure this same state to contain a thermal flux of her
$t$-particles.

In general relativity there are more possibilities. Since the
theory is generally covariant, any time coordinate, possibly defined only
locally within a
patch, is a legitimate choice with which to define particles.  Of course in
a given
spacetime, there may be particular choices for coordinates that are more
interesting than
others from the point of view of physical interpretation. For example, far
from a star
spacetime becomes flat, and asymptotically inertial Minkowski
coordinates $(t, x^i )$ are useful.  Suppose now that the star collapses to form a black
hole.  Far from the black hole, spacetime is still asymptotically flat.
Consider a wave packet which starts far from the star and
propagates through the collapsing star,
such that it just escapes being captures by the forming black hole and
propagates back out
to the flat region.
Suppose that the wave starts out composed only of positive frequency
waves with respect to the time coordinate in the asymptotic region $t$.
When the packet
passes just outside of the forming horizon, it is in a high-curvature
region.  The field
evolves so that when it is again far from the black hole, it will be a
mixture of positive
and negative frequency components. The new, negative frequency part
corresponds to
quantum-particle production. This is the effect that Hawking calculated in
his 1975 paper
\cite{hawking}.

\vskip 0.1in\noindent
{\it Canonical Quantization, Hilbert Space and Particle Number Operators}
\vskip 0.05in

Next we sketch the mathematical structure necessary for turning the
scenario described above into
a calculation. A reader who does not know quantum field theory will
certainly not be able to master it from the next few paragraphs.  However,
we have
tried to provide a complete enough set of definitions and relations, so
that these notes are more or less self-contained.  We will be thinking of
quantum field
theory as a linear algebra system and will ignore the problems of regulating and
renormalizing the theory to deal with infinities. Quantum operators will be
assumed to be
``normal ordered", so that their matrix elements are finite.  Complete
treatments of
quantum field theory in curved spacetime can be found in \cite{parker, bad}.

One standard way to implement canonical quantization is the following.
Choose a complete basis $\f$ of solutions to the scalar wave equation
(\ref{wave}), in the
spacetime with metric $g_{ab}$.  As a consequence of the wave equation, the
basis functions
are orthonormal $(\f ,f_{\omega '} )=\delta (\omega -\omega ' )$	with
respect to the
conserved inner product
\be\label{innprod}(f, h)=-i\int d^3x\sqrt{-g}\left( f\dot h^*-\dot f
h^*\right),\ee
where the integral is taken over a Cauchy surface and dot denotes a time
derivative.
For example, in Minkowski spacetime with metric $g_{ab} =\eta _{ab}$, the
standard choice of
basis functions for a scalar field is the set $\{ \f , \f^* \}$, where
\be\label{flatmodes}
\f = {1\over \sqrt{2\omega}} e^{-i(\omega t-\vec k\cdot\vec x)}\ee
and
$\omega = +\sqrt{\vec k\cdot\vec k}$.  The modes $\f$ are the positive
frequency modes.

The quantum field $\phi$ can be expanded in this basis as
\be\label{basis}\phi = \int d\omega (  \a \f + \a^{\dagger} \f^* ) ,\ee
where the expansion coefficients $\a$ and $\ad$ are operators.
For compactness, we are explicitly writing only the energy eigenvalue
$\omega$ and
suppressing other eigenvalue indices. The canonical commutation relations
for the scalar
field then imply commutation relations for the mode operators $\a ,\ad$,
\be\label{commute} [ a_{\omega '},\ad ] =\delta (\omega ' -\omega ),\qquad
[ a_{\omega },a_{\omega '} ]=[a^\dagger_\omega, a^\dagger_{\omega^\prime}]=0.\ee
The vacuum, or lowest energy state, which we denote $|0>_{in}$,
is the state which is annihilated by all the annhilation operators $\a$,
\be\label{avac} \a |0>_{in} =0 \quad\ee
for all $\omega>0$.
The standard Fock space of states is then constructed by applying arbitrary
products of creation operators to $|0>_{in}$.  For example,
the state $(\a ){}^n|0>_{in}$ contains $n$ {\it in}-particles of energy
$\omega$. This is
made precise by defining the number operator
\be\label{numbop}N_{\omega}^{in} =\ad \a, \ee
so that $<0| \ad {}^n (N^{in}_\omega )   \a {}^n|0> = n$.
We are calling these ``{\it in}" particles to agree with later notation.

Let us now introduce a second basis of solutions to the scalar wave
equation (\ref{wave})
$\{ \p,\p^*\}$.  The scalar field $\phi$ has an expansion in this basis as well,
\be\label{basistwo}\phi = \int d\omega (  \b \p + \b^{\dagger} \p^* ), \ee
with new creation and annhilation operators satisfing the commutation relations
\be\label{commuteb} [ b_{\omega '},\bd ] =\delta (\omega ' -\omega ),\qquad
[ b_{\omega },b_{\omega '} ]=[b^\dagger_\omega, b^\dagger_{\omega^\prime}]=0.\ee
The annhilation operators $\b$ define a second vacuum state, $|0>_{out}$,
satisfying
\be\label{bvac}\b |0>_{out} =0\ee
for all $\omega>0$.
A second Fock space of states is built from $|0>_{out}$ by applying the
creation operators $\bd$. The {\it out}-particle number operator,
$N_\omega^{out}$, measures
the number of {\it out}-particles in a state,
\be\label{numbopb}N_{\omega}^{out} =\bd \b ,\ee
so that, {\it e.g.} $<0| \bd {}^n (N_\omega ^{out} )   \b {}^n|0>_{out} = n$.

\vskip 0.1in\noindent
{\it Bogoliubov Transformations}
\vskip 0.05in

In order to calculate particle production,
we will need to express the number operator $N_{\omega}^{out}$ for the {\it
out}-particles
in terms of the creation and annhilation operators for the {\it in}-particles.
Define the linear transformations which relate one basis to the other  by
\bena\label{bogone}
\p &&= \int d\omega ' ( \alp f_{\omega '} +\bet f_{\omega '}^* )\\
\f &&  \int d\omega ' ( \alpha_{\omega ' \omega}^* p_{\omega '} -\beta_{\omega '
\omega} p_{\omega '}^* ).\eena
The coefficients in these expansions, $\alp$ and $\bet $, called the Bogolubov
coefficents, are given by the inner products
\be\label{bogtwo}\alp =(\p ,f_{\omega '} ) ,\qquad \bet =-(\p , f_{\omega
'}^* ) \ee
As a consequence of orthonormality of the basis functions, the Bogolubov
coefficients satisfy
\be\label{bognorm} \int d\omega ' (|\alp |^2 - |\bet |^2) =
\delta (\omega -\omega ')\ee
Further, we have the relation between the {\it out} and {\it in} mode operators
\be\label{bogthree} \b =\int d\omega '\left(
\alpha_{\omega\omega^\prime}^*a_{\omega^\prime}-
\beta^*_{\omega\omega^\prime}a^\dagger_{\omega^\prime}\right).\ee

We can now evaluate the expression (\ref{numbopb}) for $N_{\omega}^{out}$
in the {\it
in}-vacuum state, with the result
\be\label{outin}{}_{in}<0|  (N_\omega ^{out} )   |0>_{in} \equiv
{}_{in}<0|\bd \b |0>_{in} = \int d\omega ' | \beta  _{\omega \omega^\prime
} |^2 \ee
We see that although the {\it in}-vacuum is empty of {\it in}-particles, in
general
it will contain {\it out}-particles, because these particle states are defined
with respect to different time coordinates.

To summarize, for a particular calculation one must specify the state of
the system, here
taken to be the {\it in}-vacuum.
States and operators may be expanded in terms of different bases for the Hilbert
space. In general, a different choice of basis includes a different
choice of a time coordinate, and hence a different definition of a particle.
We work in the Heisenberg representation in which,
once specified, the state of the system is fixed and the operators evolve
in time. The
expectation values of operators/observables of the quantum field $\phi$ are
computed in the
state of the system that has been specified.

In the following we will study three examples of particle production
calculations. In each case the
strategy will be the same. We will make a choice for the state of the
system, and compute the
particle content for various observers with their various definitions of
particles.
These choices are the physics input and are determined by what questions
one wants to answer!

\section{Accelerating Observers in Flat Spacetime}

Consider an  observer in flat, Minkowski spacetime who undergoes
constant acceleration, {\it i.e.} the magnitude of his four-acceleration is
a constant.
We call this observer a Rindler observer.
The Rindler observer uses proper time along his worldline as a time coordinate.
In this example, we will compute the particle production which he observes,
and find an interesting result. The Minkowski vacuum, which is empty of
particles defined with respect to a global inertial time coordinate,
is populated by a thermal bath of particles according to particle-detectors
carried by the accelerating Rindler observer! This example is particularly instructive,
because the calculations can be done exactly (there is no scattering), and
so one clearly sees how the change of basis works.  This calculation
is in many standard texts, see {\it e.g.} \cite{bad} for a detailed pedagogical
treatment.  Our presentation will make use of a different
choice of basis functions than those usually empoyed, which will generalize more
easily to black hole spacetimes.

For notational simplicity we will work in 1+1 dimensional Minkowski
spacetime,
\be\label{flat}ds^2 = - dt^2 +dx^2 = -d\ub  d\vb \ee
where $\ub =t-x ,\vb =t+x$ are respectively ingoing and outgoing null
coordinates.
The $4$-dimensional calculation is essentially the same.
The standard quantum field theory choice for the positive
frequency modes of a massless field are the functions $\psi (x,t)\sim
e^{-i\omega (t\pm x)}$.
Rindler spacetime is the wedge region I of Minkowski spacetime, shown in
figure (2), that is covered
by the coordinate patch
\be\label{rinmetric}ds^2 =e^{2a\xi }(-dT^2 +d\xi ^2) =-e^{a(v-u)} dudv\ee
where $u=T-\xi , v=T+\xi$.  The Rindler metric (\ref{rinmetric}) is just a
coordinate
transformation of (\ref{flat}), with
\be\label{rincoord}v= {1\over a}ln\vb, \qquad  u=-{1\over a}ln (-\ub).\ee
A Rindler observer at constant spatial coordinate $\xi$ undergoes
constant acceleration with magnitude $ae^{ -a\xi }$, and the observer's proper
time coincides with the coordinate $T$.
A Rindler observer always stays within region I and
the boundaries of this wedge, along the lines $(t=\pm x)$, are
Cauchy horizons for these observers.
The $T$-translation Killing vector
${\partial\over\partial T}$
has zero norm on these horizons.  This corresponds to the fact that
${\partial\over\partial T}$ is a boost
Killing vector with respect ot the original Minkowski coordinates in
(\ref{flat}). Due to the
Cauchy horizons, the particle production calculations in Rindler and black
holes spacetimes
are very similiar.

The conformal (or Penrose) diagrams of  3+1 Minkowski, and 1+1 Minkowski with
the Rindler wedge are shown in figures (1) and (2). In general, such
diagrams are constructed
by conformally  compactifing the spacetime. The convention is that null paths
are 45 degree lines, so the causal structure can be easily read off.
See {\it e.g.} \cite{he,wald} for details.
\begin{figure}[!ht]
\begin{center}
{\epsfysize=1.75in \epsfbox{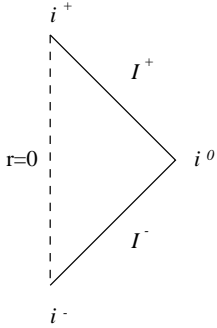}}
\end{center}
\caption{Penrose diagram for 3+1 dimensional Minkowski spacetime. The
radial-time plane is
shown, and each point is an $S^2$.
The conventions are that $\pinf$ is past null infinity, $\finf$ is future null
infinity. $i_-$ is past timelike infinity, and  $i_+$  is future timelike
infinity. $i_o$ is spacelike infinity. In figures below wavy denote curvature
singularities. The dashed line above is the  origin of radial coordinates.}
\label{f1}
\end{figure}
\vskip 0.1in\noindent
{\it Inertial and Rindler Bases}
\vskip 0.05in

To highlight the similiarities, we will call the Rindler horizons $\ph$
and $\fh$, in analogy with an eternal black hole.
First, let's define the relevant  positive and negative
frequency modes, and then turn to the issue of normalization. One basis
for the space of solutions to the scalar wave equation (\ref{wave}) in the
global inertial
coordinates (\ref{flat}) are the functions $\f$, $\f{}^*$, $\j$ and
$\j{}^*$ given by
\be\label{minkmode}\f ={1\over \sqrt{2\pi \omega} }e^{-i\omega \vb }, \qquad
\j = {1\over \sqrt{2\pi \omega} }e^{-i\omega \ub } .\ee
The functions $\f$ are positive frequency inward propagating modes, while
the functions
$\j$ give positive frequency outward propagating modes. These modes are
normalized with respect to the inner product (\ref{innprod}) .
One expansion for the scalar field $\phi$ is then
\be\label{rinphi}\phi = \int d\omega (  \a \f + \a^{\dagger} \f{}^*  +
\d \j + \d^{\dagger} \j{}^* ) .\ee
The mode operators
$\a$ and $\d$ are taken to annhilate the global inertial vacuum $|0>$
\be\label{advac} \a |0>= \d |0> =0\ee
for all $\omega>0$.  We will take $|0>$ to be the quantum state of the scalar
field.
\begin{figure}[!ht]
\begin{center}
{\epsfysize=1.75in
\epsfbox{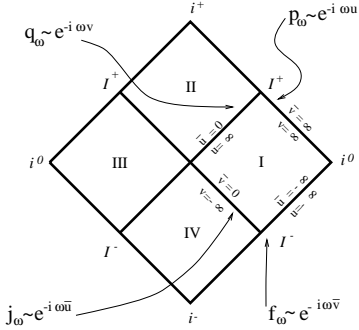}
}
\end{center}
\caption{Penrose diagram for $1+1$ dimensional Minkowski spacetime.
Minkowski lightcone coordinates are $(\bar u,\bar v)$.
Region I is the wedge covered by the
Rindler coordinates $(u,v)$. There is a symmetrical wedge on the left hand
side; the
calculations below are done in region I. The modes which define positive
frequency on each of
the boundaries of region I are indicated.}
\label{f2}
\end{figure}

For the Rindler observer, we define the modes $\q$, $\q{}^*$,
$\p$ and $\p{}^*$ given by
\be\label{rinmode} \q ={1\over \sqrt{2\pi \omega} }e^{-i\omega v } ,\qquad
\p = {1\over \sqrt{2\pi \omega} }e^{-i\omega u } ,\ee
which are only defined in the Rindler wedge.
A second expansion for the scalar field $\phi$ is then
\be\label{rinphitwo}\phi = \int d\omega ( \b \p + \b^{\dagger} \p{}^*  +
\c \q + \c^{\dagger} \q{}^* ). \ee
The Rindler mode operators $\bd$ and $\cd$ are creation operators for
inward and outward propagating Rindler particles respectively. The number
of particles that
the accelerating observer measures near $\finf$ is then given by
(\ref{outin}),  where the
in-vacuum vacuum is  defined with respect the the global inertial time
coordinate, as in
(\ref{advac}).

\vskip 0.1in\noindent
{\it Normalized wave packets}
\vskip 0.05in

The mode functions are normalized in the sense of distributions,
however each mode is not square integrable. To get a finite result for a
particle production calculation, one needs to form square-integrable wave
packets.
Let $F_{\omega}(\ub ,\vb )$ be the solution to the wave equation which is
equal to  a specified positive frequency wave packet on $\pinf$,
\be\label{packet}F_{\omega}\rightarrow \int d\nu W(\nu -\omega )f_{\nu}
(\vb ),\qquad
\vb \rightarrow \pinf.\ee
Here $W(x)$ is a ``window function" that is
peaked about the origin and chosen such that the packet is peaked
about $\vb$ near $\pinf$. In particular, the function $F_{\omega}$ vanishes
on $\fh$.
Rather than introducing corresponding new notation for all the modes, we
will indicate the
places in the calculations where it is necessary to sum up the plane wave
modes to make
normalizable states $F,J,Q,P$.

In the black hole calculation, boundary conditions on the
scalar field $\phi$ are set on $\pinf$, so we will proceed analogously here.
Given a positive frequency, outward propagating Rindler wave packet on $\finf$,
one wants to solve the wave equation to find the form of the wave packet in
the far past.
One then decomposes this into a sum over positive and negative frequency
parts with respect
to the Rindler coordiante $v$. The calculation is most simply carried out
mode by mode, {\it i.e.} for $\phi \rightarrow  e^{-i\omega u}$ on $\finf$.
In the
Rindler wedge, the past boundary is $\ph$ plus $\pinf$. Because spacetime
is flat, there is
no scattering of the scalar field $\phi$. Therefore, the wave packet above
propagates from $\finf$ to $\ph$, and none reaches $\pinf$.  The particle
production comes
solely from the change of basis, {\it i.e.} from different definitions of time.

\vskip 0.1in\noindent
{\it Particle Production}
\vskip 0.05in

To compute the flux
of Rindler particles across $\finf$ we only need the Bogoliubov coefficients
as in (\ref{bogtwo})
\be\label{rinbog} \alp = (\p ,j_{\omega '} )_{\ph}  ,\qquad
\bet=  -i\alpha_{\omega ,-\omega '},\ee
where the first integral is taken over the past Cauchy horizon $\ph$.
The mode functions satisfy  $\partial_{\ub}\p{}^* =(i\omega / a \ub)\p $,
so that
\bena\label{rinalp}\alp &=& {-1\over 4\pi \sqrt{\omega \omega '}}
\int_{-\infty}^0 d\ub (\omega ' -{\omega \over a\ub } ) e^{i\omega ' \ub}
e ^{i {\omega \over a} ln(-\ub )} \\
&=&{i\over 2\pi }{1\over \sqrt{\omega ' \omega }}
(i\omega ' )^{-i {\omega \over a}}\Gamma (1 +i {\omega \over a} ), \eena
where $\Gamma (s) =\int _0 ^{\infty} e ^{-z} z^{s-1} dz$, and
we have used $\Gamma (1+s) =s\Gamma (s)$.
This is the same expression that we will find for the Bogoliubov coefficients
$\alp$ in the black hole case, with the acceleration $a$ being replaced by
the surface
gravity of the black hole. With a bit more analysis, which we defer until
the black
holes calculation, we will find after restoring factors of Planck's
constant $\hbar$ the
result for the number of particles produced in each Rindler mode
\be\label{rinspec} <N_{\omega }^{rind} > = {1\over e^{ {2\pi \omega \over
\hbar a} } -1 } ,\ee
which is a black body or thermal spectrum, with temperature
\be\label{rintemp}T=\hbar {a\over 2\pi}.\ee

\section{Black Holes}
A stationary black hole spacetime\begin{footnote}
{This will not be a comprehensive introduction
to black holes! See, {\it e.g.} \cite{wald} for details, proofs, and
further properties.}
\end{footnote}
has a
killing vector $\xi^a$ which is normal to the horizon, and whose norm $\xi
^a \xi _a =0$ on the
horizon. The surface gravity $\kappa$ is defined by
$\nabla ^b (\xi ^a \xi _a )=-2\kappa \xi ^b$ on the horizon. The horizon area
$A$ is the area of the intersection of the horizon with a constant time
slice, which is a
two-sphere in all of the cases considered here.

According to Birkhoff's theorem, the Schwarzschild metric below is the
unique spherically
symmetric solution to the vacuum Einstein equation $R_{ab}=0$,
\be\label{schw}ds^2  = -V(r)dt^2 +{dr^2 \over V} +r^2
d\Omega ^2,\qquad V(r) = 1-{2M\over r}.\ee
Here $d\Omega ^2$ is the volume element on the unit $2$-sphere.  The
spacetime has a black
hole horizon where the norm of the time-translation
killing vector ${\partial \over \partial t}$ vanishes.  In
the  coordinates (\ref{schw}) the horizon is at $r=2M$ and has area $A=4\pi
M^2$.
The parameter $M$ is the ADM mass of the spacetime.  For any static black
hole with
metric of the form (\ref{schw}), possibly with a different function $V(r)$,
the surface
gravity is given by $\kappa ={1\over 2} V'(r_H )$, where $r_H$ is the
horizon radius. The
metric (\ref{schw})  has a coordinate singularity at $r=2M$ and a curvature
singularity at
$r=0$.
\begin{figure}[!ht]
\begin{center}
{\epsfysize=1.75in \epsfbox{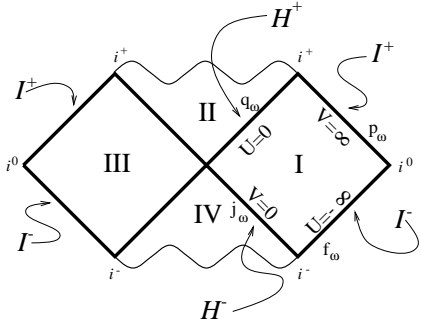}}
\end{center}
\caption{Penrose diagram for an eternal Black Hole (Extended Schwarzchild):
Regions I and III are
asymptotically flat, Region II is the black hole, and Region IV is the
white hole.
For an observer in Region I,$\fh$ is the (future) black hole horizon and $\ph$
is the past black hole, or white hole, horizon. $(U,V)$ are the Kruskal
coordinates.}
\label{f3}
\end{figure}

For the particle production calculation, we will need the black hole metric
in several
different coordinate systems,
\bena\label{other} ds^2 &= &V(r)( -dt^2 + dr^{*2} ) + r^2 d\Omega ^2 \\
&=& -{2M\over r} e^{-r/2M} e^{(v-u)/4M} dudv    +r^2 d\Omega ^2 \\
& =& -{2M^3 \over r} e^{-r/2M}dU dV  +r^2 d\Omega ^2.\eena
The radial coordinate $r^*$ is known as the tortoise coordinate, $u$ and
$v$ are a
pair of ingoing and outgoing null coordinates and, finally, $U$ and $V$ are
ingoing and
outgoing null Kruskal coordinates.  The relations between the different
coordinates are
given by
\bena\label{coordtrans}
 dr^* &=& {dr\over V(r) },\quad r^*=r +2M ln ({r\over 2M}-1) \\
u &= &t-r^* ,\quad v=t+r^* \\
U &= &-e^{-u/4M} ,\quad  V=e^{v/4M} ,\eena
where factors of $r$ are understood to be implicit functions of $r*$,
$u,v$, or $U,V$ respectively. The definition of $dr^*$ is rather general,
though usually
one can't do the integral explicitly.
The black hole horizon is regular in the  Kruskal coordinates.
One finds that the Schwarzschild  coordinates (\ref{schw}) actually only
cover part of the
manifold, but that the Kruskal coordinates cover the extended  spacetime.
These features of the geometry are diplayed in the conformal (or Penrose)
diagrams, figures (2) and (3).

Finally, we will look at solutions to the scalar wave equation in the
Schwarzchild geometry.  Writing $\phi$ as the product
\be{separate}\phi _{\omega lm} (t,r*,\Omega )=\psi (r*) Y_{lm}(\Omega)
e^{-i\omega t},\ee
the wave equation (\ref{wave})  reduces to the radial equation
\be\label{waveschw} (\partial ^2 _t  -\partial ^2 _{r*} +W(r) )\psi =0,
\qquad W(r) =(1- {2M\over r}) ({2M\over r^3 } +{l(l+1)\over r^2} ).\ee
Note that in terms of the tortoise coordinate, the horizon $r=2M$ is
$r^*\rightarrow
-\infty$, whereas in the asymptotically flat limit $r\rightarrow\infty$, we
also have
$r^*\rightarrow\infty$.
In the asymptotic region $r^*\rightarrow\infty$, the potential behaves as
$W(r)\rightarrow
{l(l+1)\over r^2}$, and near the horizon $r*\rightarrow -\infty$, we have
$W(r)\rightarrow e^{r*/2M}$.  Therefore, both near infinity and near the
horizon, the
solutions $\phi _{\omega lm}$ are plane waves in $t\pm r*$, {\it i.e.}
plane waves in $u,v$.
These solutions to the wave equation will be used to define the bases for
the Hilbert
space of states. For notational simplicity, as in section (2), we will
supress the $l,m$
subscripts in the following calculations.

\section{Particle Emission from Black Holes}

Our strategy will be to first present Hawking's original calculation of
particle production,
which is done in a gravitational collapse spacetime.  In the following
sections, we will then
compute the same result for the eternal black hole, as well as an analogous
result for a
charged eternal black hole in a  spacetime which is asymptotically deSitter.

\vskip 0.1in\noindent
{\it Defining the Problem}
\vskip 0.05in
First let us outline the idea.
Hawking \cite{hawking}  originally did the calculation of particle emission
for a black hole that is formed by gravitational collapse. In
the far past, the spacetime is nearly Minkowski, the largest gravitational
effects being at the surface of the star, and we can assume that the
quantum state is empty of {\it in}-particles near $\pinf$.
We will call this state $\vac$.
The star collapses to form a black hole.  Hawking found that near $\finf$,
the state $\vac$
contains a thermal flux of {\it out}-particles. The particles produced are
known as Hawking
radiation.
\begin{figure}[!ht]
\begin{center}
{\epsfysize=1.75in \epsfbox{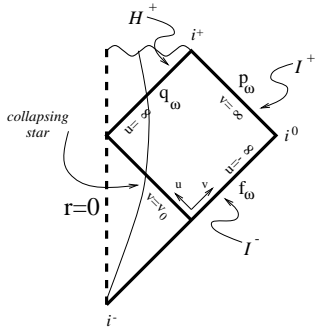}}
\end{center}
\caption{Penrose diagram for a black hole formed via gravitational
collapse: The boundary of
the collapsing star is shown. The star interior covers up
regions III and IV of the extended black hole spacetime. Spacetime
curvature is small inside the star.
At some point during collapse, the star falls within its event horizon,
and the black hole forms.}
\label{f4}
\end{figure}

There is no white hole horizon in the collapse spacetime,  since
$\ph $ is replaced by the interior of the collapsing star. From the
conformal diagram in figure (4), one sees that $\pinf$ is a Cauchy surface.
In order to choose a set of basis functions that define particle states in the
far past, one must choose a time coordinate with which to define positive
frequency oscillations on $\pinf$.  We will take the the early time positive
frequency modes to be the solutions $\f$ to the wave equation that
behave near $\pinf$ like
\be\label{inmodes} \f (u,v)\rightarrow e^{-i\omega v}\ee
Far from the star spacetime becomes flat and $v$ becomes
an ingoing null coordinate for the flat space wave equations. Therefore,
this choice of
positive frequency modes corresponds to the usual Minkowski particle
states. Note that
$v$ is the affine parameter for the null geodesic generators of $\pinf$.

Define creation and annihilation operators $\ad$, $\a$ for these, as in
(\ref{basis}), via
the  expansion
\be\label{basthree}\phi(u,v) = \int d\omega (  \a \f + \a^{\dagger} \f^* ) \ee
The vacuum is then taken to satisfy
\be\label{vacstate} \a \vac =0, \ee
for all $\omega>0$.
Note that this state is annhilated by the $\a$ at all times. The label
{\it in} refers to the fact that the
boundary conditions on the modes $\f$ are fixed on $\pinf$.

In order to define a complete set of particle states at late times, we must
define modes on both $\finf$ and $\fh$, because $\finf$ itself is not a Cauchy
surface. On $\finf$ we take the {\it out}-states to be solutions
to the wave equation with boundary conditions that on $\finf$
\be\label{outmodes}\p \rightarrow e^{-i\omega u}.\ee
The coordinate $u$ is an outgoing null coordinate and is the affine parameter
for the null geodesic generators of $\finf$.
Again, this choice of positive frequency late time modes coincides
with the usual choice in Minkowski spacetime.

In order to form a complete basis, we must add modes which define particle
states on $\fh$
and it's extension through the collapsing matter. Here we cannot
make a choice based on a flat spacetime limit. One
approach to this problem is as follows \cite{hawking}.  Choose any set of modes
$\q$ that are well behaved on $\fh$. The choice of quantum state $\vac$
implies that at early times, the density matrix\begin{footnote}{The density
matrix
formulation of quantum mechanics is a generalization of the standard
Schroedinger/Heisenberg wave mechanics, which is needed for quantum
statistical mechanics.}\end{footnote}
of the system is simply
\be\label{density}\rho =\vac {}_{in} <0| ,\ee
the density matrix for the ``pure state" $\vac$. The operator
$\rho$ can be expanded in either the {\it in} or {\it out} basis. Expanding
$\rho$ in the
$\f ,\q $ basis, it is a product of the ``$\fh$" Fock space, constructed with
the mode operators $\cd$ and the ``$\finf$" Fock space, constructed with
the operators
$\bd$.
The expectation value of any operator $O^{AF}$ that only depends
on the degrees of freedom in the asymptotically flat region of the
spacetime (region I in Fig 3) may be computed using the reduced density
matrix $\rho ^{red} \equiv Tr_{\{q\}}\,\rho$ as
\be\label{reduced}<O^{AF}>= Tr (\rho^{red} O^{AF} ) \ee
The reduced density matrix,
$\rho ^{red}$, is the same for all bases related by unitary transformations
to the chosen basis. Therefore $<O^{AF}>$ is independent of the choice of modes
$\q$ on the black hole horizon $\fh$.

Therefore, the scalar field $\phi$ can also be expanded in the {\it out}-basis,
\be\label{basfour}\phi = \int d\omega (  \b \p + \b^{\dagger} \p *
\c \q + \cd \q * )\ee
As discussed in the Rindler example,  one must use normalized
wave packets to have a finite result for the number of particles
produced in given a frequency interval, per unit time.  Again, here we will
do the calculation individually for each eigenmode and assemble wave
packets at the end.

Of course, solving the wave equation in the black hole spacetime is
harder than in the Minkowski case!  In the black hole case, we don't know
global analytic
solutions. Consider a wave packet peaked about frequency $\omega$ that
propagates inward
from $\pinf$ towards the horizon of
an eternal black hole. Roughly speaking, the wave
scatters in two parts. A fraction $1-\Gamma _{\omega}$ of the packet
backscatters
off the  curved geometry, {\it i.e.} due to the potential $W(r)$ in
(\ref{waveschw}), and propagates
out to $\finf$, essentially without a change of frequency.
The remaining fraction $\Gamma _{\omega}$ propagates parallel to $\ph$ and
is absorbed by the black hole horizon. It is this second portion
that leads to the particle production.  Therefore, we can write
$\f=\f{}^{(1)}+\f{}^{(2)}$,
where the superscripts $(1)$ and $(2)$ denote these two parts, and
similarly for the
functions $\j$, $\p$ and $\q$.  We can also write for the Bogoliubov
coefficients and the
scattering coefficient $\Gamma_\omega$,
\bena\label{parts}\alp &=&\alp{}^{(1)} \delta_{\omega ' \omega}
+\alp{}^{(2)} ,\qquad
\bet =\bet {}^{(2)} \\
\Gamma_{\omega} &= &\int d\omega ' (|\alp {}^{(2)} |^2 - |\bet {}^{(2)}
|^2) \eena
To compute particle production, one can ignore the backscattered $(1)$
component of the wave.  In addition, for simplicity we will drop the
superscript $(2)$
on the coefficients in the following.

Hawking did the calculation by studying a wave propagating
backwards in time in the collapsing star spacetime.  Choose
boundary conditions such that the wave is positive frequency on $\finf$,
so that the scalar field $\phi\rightarrow\p$ as in (\ref{outmodes}).
The goal is then to solve for the behavior of the scalar field $\phi$ on
$\pinf$ and to decompose the wave into positive and negative frequency parts there.
Given this choice of boundary conditions, the wave propagates backwards in
time, and the collapsing star geometry sets the natural definitions
of positive and negative frequency in the far past.

In section (6), we will compute black hole radiation in the extended, eternal
Schwarzschild spacetime with a particular choice of positive frequency on $\ph$.
This choice is dictated by the results in the collapsing black hole calculation.
Importantly,
learning what choice to make for positive frequency modes on
horizons allows one to extend the calculation of Hawking radiation to black
holes
in spacetimes that are not asymptotically flat \cite{ktrnds}.  We will outine
one such calculation in section (6).

\vskip 0.1in\noindent
{\it The Calculation}
\vskip 0.05in

\noindent
1) The mode (\ref{outmodes})  propagates along a path $\gamma$ that
goes from $\finf$ along a geodesic $u=u_1$
passing close to the black hole horizon $\fh$. The ray passes through the
collapsing star and then propagates out to $\pinf$ along a geodesic $v=v_1$,
which is close to $v=v_0$. The ray $v=v_0$, shown in figure (4), is the
last inward
propagating ray on the surface of the star that reaches $\finf$. Inward
propagating rays
with  $v>v_o$ enter the black hole. In the extended spacetime $v=v_o$ would be
the white hole horizon $\ph$.

\noindent
2) The ray $\gamma$ is connected to $\fh$ and $v=v_o$ by
a geodesic deviation vector $\epsilon n^a$ with $\epsilon $ small and
positive. On the
part of the path that passes close to $\fh$, $n^a$ is tangent to a null
geodesic which is
ingoing at $\fh$. The normalization is fixed by the condition $n^a l_a
=-1$, where
$l^a$ is a null geodesic generator of $\fh$.

\noindent
3) Let $p^a$ be tangent to an ingoing null geodesic at $\fh$,
$p^a = {du\over d\lambda}{\partial \over \partial u}$. Note that $p^a$ is
parallel to $n^a$ and therefore satisfies
\be\label{parallel}p^b=A^2 n^b.\ee
Solving the geodesic equation for $p^a$ near $\fh$ gives the affine parameter
$\lambda$ in terms of the coordinate $u$,
\be\label{affine}\lambda =-B^2 e^{-\kappa u} =B^2 U \ee
where $\kappa$ is the surface gravity of the black hole, and $U$ is the
Kruskal coordinate defined in (\ref{coordtrans}).  This expression will be
useful below.
For the Schwarzchild case, $\kappa =1/4M$, but the relation (\ref{affine})
will generalize to
other cases as well. The affine parameter $\lambda =0$ on $\fh$.

\noindent
4) The affine parameter is a good coordinate near $\fh$, while $u$ is not.
So the
deviation vector connects the two null geodesics, the horizon at $\lambda
=0$ and
the ray $\gamma$ at $\lambda$, where $\lambda <0$.

\noindent
5) In these local inertial coordinates, the geodesic equation is simply
${dp^\mu \over d\lambda} ={d^2 x^\mu\over d\lambda ^2}=0$, so that
\be\label{difference} \lambda p^\mu =x^\mu (\lambda ) -x^\mu (0) =-\epsilon
n^\mu,\ee
where the last equality follows from the definition of the deviation
vector in point (2) above.  Equations (\ref{difference})  and
(\ref{parallel}) then imply that
\be\label{eps}\epsilon =-\lambda A^2\ee

\noindent
6) Next, we trace the ray $\gamma$ through the collapsing star, and back
to $\pinf$. In the conformal diagram, the ray bounces off the origin
of coordinates and follows a null geodesic of constant $v <v_0$, which is
near $v=v_0$. The two geodesics are still connected by $\epsilon n^a$. Since
spacetime is approximately flat on this part of $\gamma$, we
have
\be\label{changev}v_0 -v =\epsilon =-\lambda A^2 =C^2 e^{-\kappa u} ,\ee
which holds\begin{footnote}{On the conformal diagram, the deviation vector
appears to flip direction when it ``turns the corner".  This is simply  because
the vertical left hand boundary is the origin of coordinates, so that the
rays are
reflected rather than continued.  Note that the signs are consistent
through the chain of equalities in (\ref{changev}).  I would like to thank the
students in my $1999$ General Relativity class for patiently helping to sort out
the signs.}\end{footnote}
on $\pinf$.

Equation (\ref{changev}) is the desired relation between $u$ and $v$.
For a solution to the scalar wave equation $\nabla _a \nabla ^a \phi =0$ that
has the boundary condition $\phi\sim e^{-i\omega u}$ at $\finf$, we have
on $\pinf$
\bena\label{inwave}
\phi &\sim &
e^{i{\omega \over\kappa}ln ({v_0 -v \over C^2})},\qquad    v<v_0 \\
\phi &\sim & 0,\qquad\qquad\qquad\   v>v_0 .
\eena
The wave vanishes for $v>v_0$ because it would have had to come out
of the black hole horizon to reach this part of $\pinf$.  Proceeding as
before, one finds
the expressions for the Bogoliubov coefficients
\bena\label{bhalp}\alp &=& (\p ,f_{\omega '} )_{\pinf} = {1\over
2\pi\sqrt{\omega\omega^\prime}}
\int_{-\infty}^0 dv \left(\omega ' -{\omega \over \kappa v }\right)
e^{i\omega ' v}
e ^{i {\omega \over \kappa} ln( -v  )}  \\
&= &{1\over i\pi \sqrt{\omega \omega '}}
(i\omega ' )^{-i {\omega \over \kappa}}\Gamma (1 +i {\omega \over \kappa} ) \\
\bet &=&-i\alpha _{\omega ,-\omega '},\eena
where we have set $v_0=0$ in the above expressions.

The Bogoliubov coefficient
$\alp$ is analytic in the lower half of the complex $\omega '$ plane, because
it is the fourier transform of a function which vanishes for $v>0$.  The
coefficient
$\alp$ has a logarithmic branch point at $\omega '= 0$, so the branch cut
extends into the upper half plane. Therefore, we have
\be\label{abratio}|\alp |=e^{\pi\omega /\kappa} |\bet |.\ee
The spectrum of produced particles that then follows, making use of
(\ref{outin}) and
(\ref{parts}),
is given by
\bena\label{bhspec} <N_{\omega }^{bh} > & =&\int d\omega ' |\bet |^2 \\
 &=&{\Gamma_{\omega} \over e^{ {2\pi \omega \over
\hbar \kappa} } -1 }\eena
This is a black body or thermal spectrum, with temperature
\be\label{bhtemp}T=\hbar {\kappa\over 2\pi},\ee
with $\kappa =1/ 4\pi$  for Schwarzschild.

The coefficient $\Gamma$ entered in our discussion of normalized wave packets,
as the portion of the wave which propagates close to the horizon, through
the colllapsing star, and back out to $\pinf$. This is almost identical
to the fraction which would propagate into the white hole horizon $\ph$
if we were working in the extended spacetime, rather than the gravitational
collapse case. But this in turn is equal to the fraction of a wave
which is absorbed by the black hole horizon $\fh$ for a wave which starts
at $\pinf$. So $\Gamma _{\omega}$ is just the classical absorption coefficient
for scattering a classical scalar field off a black hole.  Direct
calculation gives
\be\label{absorption}\Gamma _{\omega}\rightarrow 1 ,\  \omega M\gg 1, \quad
\Gamma _{\omega}\rightarrow {A\over 4\pi} \omega ^2 ,\  \omega M\ll 1 .\ee
The large energy limit is just the particle
limit, in which everything is absorped.

One fascinating implication is that the classical black hole mechanics theorems
and the laws of thermodynamics have more than a formal analogy. A black hole
radiates with  temperature $T=\hbar {\kappa\over 2\pi}$, and has an entropy
\be\label{bhentropy} S_{bh}= {1\over 4} A!\ee

\vskip 0.1in\noindent
{\it Generality and Back Reaction }
\vskip 0.05in

Hawking also calculated particle production in quantum fields by charged
and rotating black holes. Calculations have also been done for emission
of fermions and gravitons, linearized perturbations of the metric.  In all
of these
cases one finds a thermal spectrum,
\be\label{thermal}<N_\omega^{bh} > = {\Gamma_{\omega} \over e^{ {2\pi
(\omega -\mu ) \over
\hbar \kappa} }\pm 1},\ee
where the +1 corresponds to fermions and -1 to bosons. In thermodynamics,
$\mu$ is  called a chemical potential. For black hole emission, $\mu$ is such
that a charged black hole preferentially emits charged, massless particles
of the same sign as its own charge. Rotating black holes preferentially emit
particles with the same sense of angular momentum. Hence black holes
can spindown via Hawking radiation and also discharge, if there are fields
which carry the same kind of charge as the black hole.
Another generalization of interest is to black branes in higher dimensions,
which are important in string theory and will be discussed briefly below in
section (7).

In the preceeding calculation, the spacetime metric was fixed. Even though
we don't have a quantum theory of gravity to determine how
the metric evolves with the quantum particle emission, it is assumed that
the mass of the black hole decreases.  For a neutral black hole, the temperature
increases as the mass decreases, so the rate of black hole evaporation
increases with time.  Very small black holes have very large curvatures,
and at some point the classical gravity description is not valid. So
the endpoint of this run away evaporation is not something we can compute
and has been the
subject of much debate.

However, the situation is rather different for particle production
from charged black holes. A static, spherically symmetric, charged black is a
solution to the Einstein-Maxwell equations, {\it i.e.} equation
(\ref{einstein}) with $T_{ab}$ given by the
stress-energy of the Maxwell field.
We will take the black hole
charge $Q$ to be positive.  The Reissner-Nordstrom spacetime for an
electrically charged
black hole is given by
\bena\label{rn} ds^2  &= &-V(r)dt^2 +{dr^2 \over V} +r^2 d\Omega ^2,
\quad V(r) = 1-{2M\over r} +{Q^2 \over r^2}  \\
A_b dx^b &=&-{Q\over r}dt .\eena
Here $A_b$ is the $U(1)$ electromagnetic gauge potential. The spacetime
(\ref{rn})  describes
a black hole, {\it i.e.} there is a horizon, when $M\ge Q$.  For $M<Q$
there is no
horizon and the spacetime has a naked singularity. The case $M=Q$
is called an extremal black hole.

For the Reissner-Nordstrom black holes, the temperature is still given by
$T=\hbar
{\kappa\over 2\pi}$ with the surface gravity $\kappa$ calculated from the
metric (\ref{rn}).
For $M\gg Q$, the
temperature reduces to the Schwarzchild result.  However, as $M\rightarrow
Q $ the surface
gravity  $\kappa\rightarrow 0$, with $\kappa =0$ for $M=Q$.
Therefore, the temperature vanishes for an extremal black hole. So, for
charged black holes, if we assume that
there are no charged fields present to discharge the hole, then the
semiclassical calculation says
that a black hole with $M>Q$ evaporates down to $M=Q$, at which point the
evaporation stops.
We will return to this picture in connection with the positive mass
theorems for black
holes, and quantum mechanical ground states in  string theory in section (7).

\section{Extended Schwarzchild and Reissner-Nordstrom deSitter Spacetimes}

\vskip 0.1in\noindent
{\it Extended Schwarzchild}
\vskip 0.05in

In the extended Schwarzchild spacetime, also known as the eternal black hole,
shown in figure (3), one basis
consists of the modes $\{\f,\j\}$ with boundary conditions specified on
$\pinf$ and $\ph$ respectively.  A
second basis consists of the modes $\{ \p ,\q \}$ with boundary conditions
specified on
$\finf$ and $\fh$ respectively.  On $\pinf$ and $\finf$ we choose the
same modes as before, (\ref{inmodes}) and (\ref{outmodes}).

On the black hole and white hole horizons, we will define positive
frequency modes so that the resulting particle production is the same as
in the collapse spacetime.  Indeed, equation (\ref{affine}) implies that
the correct
choice on the horizons is to use the null Kruskal coordinates $(U,V)$ defined
in (\ref{coordtrans}).  Note that the coordinates $U,V$ are affine
parameters for the null
geodesic generators of the horizons, so this choice is consistent with the
choices of $(u,v)$ at null infinity.  We then have
\bena\label{jpmodes} \j &\rightarrow &{1\over \sqrt{2\omega}}e^{-i\omega U},
\quad {\rm near} \  \ph \\
\q &\rightarrow &{1\over \sqrt{2\omega}}e^{-i\omega V}, \quad {\rm near} \
\fh \eena
To find the Bogoliubov coefficients $\alp$, the computation in
(\ref{bhalp})  is replaced by
an integral over $\ph$, as was done for the Rindler spacetime in
(\ref{rinbog}).
The integral is then the same as in equations (\ref{rinalp})  and
(\ref{bhalp}), and the thermal
spectrum follows as before.

\vskip 0.1in\noindent
{\it Charged Black Holes in DeSitter}
\vskip 0.05in

A deSitter spacetime is a spacetime of constant positive scalar curvature,
and is a solution to the Einstein equation with cosmological constant
$\Lambda >0$, {\it
i.e.} $G_{ab}=8\pi \Lambda g_{ab}$.  A particular slicing of deSitter
describes the
Inflationary Universe. A Reissner-Nordstrom-deSitter, or RNdS, spacetime
describes an eternal charged
black hole in a spacetime which is asymptotically deSitter, rather than
asymptotically flat.
The metric and gauge field are given by the expressions in (\ref{rn}), but
with the
radial function $V(r)$ given by
\be\label{rnds}
V(r)=1-{2M\over r} +{Q^2\over r^2} -{1\over 3}\Lambda ^2 r^2. \ee
For a range of values of $Q$ and $M$, the spacetime has three Killing horizons;
inner and outer black hole horizons and a Cauchy horizon, called the
deSitter horizon.
This implies that there are two sources of particle production in an RNdS
spacetime, the
black hole horizon and the deSitter horizon \cite{gh}.
One interesting question that we will address below is whether these two
sources can ever be
in a state of thermal equilibrium \cite{ktrnds}.
\begin{figure}[!ht]
\begin{center}
{\epsfysize=1.75in \epsfbox{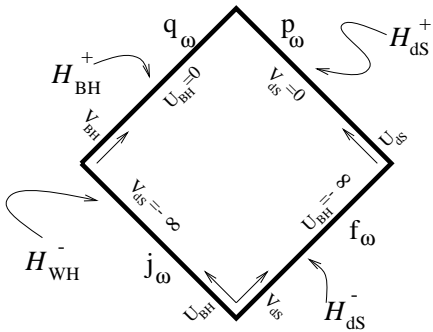}}
\end{center}
\caption{A part of the conformal diagram for RNdS (a charged black hole in
asymptotically
deSitter spacetime). The black hole, white hole, past and future deSitter
horizons
are indicated.}
\label{f5}
\end{figure}

The conformal diagram for the
relevant portion of RNdS is shown in figure (5). The region is bounded
by the white hole, black hole, past and future deSitter horizons.
Following the discussion for the extended Schwarzchild spacetime, we
define positive frequency on each of these horizons by a Kruskal-type
coordinate, {\it i.e.} a coordinate which is an affine parameter for the null
geodesic generators of that horizon.  Explicity, letting $u=t+r^*$ and
$v=t-r^*$, the
Kruskal coordinates are given by
\bena\label{kruskal}
\ubh =-{1\over {\kbh}}e^{-\kbh u}   ,&&\qquad V_{bh} ={1\over {\kbh}}
e^{\kbh v}\\
\uds={1\over \kds}e^{\kds u} ,&&\qquad V_{dS} =-{1\over \kds}e^{-\kds v} \eena
Near the black hole horizon, the metric is then well behaved and has the
limitting
form
\be\label{near}
ds^2 \approx \kbh d\ubh dV_{bh},\ee
Similarly, one can show that the coordinates $(\uds,\vds)$ are also good
near the deSitter horizon.

The Klein-Gordon equation for $\phi$ near any of the horizons reduces to
the free wave equation.  As in the case $\Lambda=0$,
the potential $W(r)$ due to the background gravitational field decays
exponentially near a horizon.  Consider then
a pure positive frequency, outgoing wave near the deSitter horizon
at late  time, $\p\sim e^{-i\omega \uds}$. In the geometrics optics
limit, finding the form of this wave
propagated back to the white hole horizon reduces to finding the
dependence of the coordinate $\uds$ on the coordinate $\ubh$.
Using the expressions in (\ref{kruskal}) , it follows that on the white
hole horizon
the quantity $G_{\omega}(\ubh )\equiv \pw (\uds (\ubh ))$ behaves as
\bena\label{whitehole}
G_{\omega}(\ubh )&\sim&
e^{-i\omega \xi ^2 ({-1\over \ubh})^{\eta}} ,\qquad
\ubh <0\\
G_{\omega}(\ubh )&\sim &0, \qquad\qquad\qquad\quad \ubh >0 .
\eena
where $\eta \equiv \kds /\kbh$ and $\xi ^2 \equiv {1\over \kds }
({1\over \kbh })^{\eta} $.
The Bogoliubov coefficients are then given by
\be\label{betabh}
\bwwpbh  = {1\over  \sqrt{2\pi\omega}}
\int d\ubh e^{-i\omega^\prime \ubh} G_{\omega} (\ubh )  .\ee

Similiarly, there is emission ``from'' the deSitter horizon as seen
by an observer outside the black hole horizon at late times.  Consider
a positive frequency wave which is entering the black hole horizon, $\q
\sim e^{-i\omega
V_{bh}}$. In the geometrics optics approximation on the past deSitter
horizon, the
quantity $F_{\omega}(V_{dS}) \equiv \q (V_{bh} (V_{dS}) $ is given by
\bena\label{desitter}
F_{\omega}(V_{dS}) & \sim & {1\over \sqrt{2\pi \omega}}
e^{-i\omega\mu ^2 ({-1\over V_{dS}})^{{1\over \eta}}},\qquad  V_{dS} < 0\\
F_{\omega}(V_{dS}) & \sim & 0, \qquad\qquad\qquad\qquad\qquad V_{dS} >0,
\eena
where $\mu ^2 =1/\kbh (1/\kds )^{{1\over \eta}}$.
Similarly to equation (\ref{betabh}), the Bogoliubov coefficients $\bwwpds$
are given in terms
of the fourier transform  of (\ref{desitter}).
For general values of $Q$ and $M$, the functions $F_\omega$ and $G_\omega$
appearing in (\ref{whitehole}) and (\ref{desitter}) are related according to
\be\label{relation}
G_{\omega}(x) =F_{{\omega\over\eta ^2}}
(x^{\eta ^2}).  \ee
We see that the two functions are equal for
$\eta =1$, which occurs when $|Q|=M$.  Therefore
$\bwwpbh =\bwwpds$ if and only if $|Q|=M$.
This implies that for each horizon, the flux of particles absorbed is equal
to the
flux of particles emited.

The spectrum of emitted particles is given by $\nw =\int d\omega '
|\bwwp |^2 $. We can estimate the above integrals using the stationary
phase approximation. It is simpler to work with the coefficients
$\awwp =-i\beta _{\omega ,-\omega '}$.
For the case $|Q|=M$, when the surface gravities or temperatures are equal,
we have
\bena\label{rndsalp}\alp &= &{-1\over 2\pi \sqrt{\omega \omega '}}
\int_{-\infty}^0 d\ubh (\omega ' +{\omega \over \kappa^2 \ubh {}^2 } )
 e^{i\omega' \ubh} e ^{i {\omega \over \kappa ^2 \ubh} } \\
&=&{-1\over 2\pi \kappa}{\omega ' \over \omega }\int_{-\infty}^0 dz
(1 +{1\over z^2}) e^{i(z+1/z)\sqrt{\omega\omega ' }/\kappa}. \eena
For large $\omega '$ the stationary
phase approximation gives
\be\label{cando}
\alp  \approx {-1\over 2\pi\kappa} {\omega '\over \omega }
e^{-{2i\over \kappa}\sqrt{\omega\omega '} }.\ee
As before, the Bogoliubov coefficients $\bet$ are obtained
by analytically continuation. Noting that (\ref{whitehole}) implies that $\alp$
is analytic in the lower half $\omega '$ plane, we have
\be\label{rndsratio} |\alp |^2 =|\bet |^2 e^{{4\over
\kappa}\sqrt{\omega\omega '} } \ee
Then (\ref{bognorm}) implies
\be\label{betbet}\beta _{\omega\nu}\beta^*_{\omega ' \nu}  =
{e^{-i\nu(\omega -\omega ') }\over
e^{{2\over \kappa} (\sqrt{\omega\nu} +\sqrt{\omega '\nu})} -1 },\ee
and finally we obtain the spectrum
\bena\label{rndsspec}<N_\omega >&=&\int d\nu \beta _{\omega\nu}\beta^*_{\omega
\nu} = \int _c^{\infty}d\nu ( e^{{4\over \kappa} \sqrt{\omega\nu} } -1 )^{-1} \\
&= &{\pi^2\over 6} ({\kappa ^2 \over 8\omega }+ {\kappa \over
2}\sqrt{c\over\omega} )
e^{-{4\over \kappa}\sqrt{c\omega})} .\eena
The form of the spectrum depends on the infrared cutoff of the range of
integration over frequencies. The integral converges if $c=0$. However,
one would expect that only wavelengths that are less than, or of order
the deSitter horizon scale should be included, {\it i.e.} $c\approx
A^{-1/2}_{dS}$.
Note that the spectrum then is not a thermal black body spectrum, though
the system is still
in an equilibrium state.

There are several limits one can take in order to check this result. Letting
$\kappa \rightarrow 0$ above, corresponds to keeping
$|Q|=M$ and letting the cosmological constant
$\Lambda$ approach zero, so that the spacetime approaches
extremal Reissner-Nordstrom. In this limit $N_{\omega}$ goes to zero,
as it should. Secondly, one can set $Q=0$, and then
let $\Lambda\rightarrow 0$, so that the metric approaches Schwarzschild.
The particle production (\ref{betabh}) from the black hole can again
be evaluated in the stationary phase approximation. One finds that
the coefficients $\awwp$ approach those for a Schwarzschild black hole in
this limit.

\section{Black Hole Evaporation and Positive Mass Theorems}
We close  by pointing out a connection, between classical
positive mass theorems in general relativity and lowest energy states
in string theory, which is made via Hawking evaporation.
Motivated by supergravity, spinor constructions have been used to prove that for
asymptotically flat solutions to the Einstein-Maxwell equations, the ADM
mass is
always greater than or
equal to the charge of the spacetime, $M\ge |Q|$. We refer the reader to
the various papers for the full statement of the results, {\it e.g.}
\cite{witten,nester} for the case without charge or horizons and
\cite{gibhull,ghhp}  for
derivations which include charge and horizons. There are also a large number of subsequent
results that include other gauge fields, different asymptotics, or higher
dimensions, see {\it
e.g.} \cite{sorkin,gktt}.

The bound is saturated, {\it i.e.}  $M=|Q|$, if and only if the spacetime
has a super-covariantly constant spinor. The super-covariant derivative
operator\begin{footnote}{These
derivative operators arise
from supersymmetry transformations that leave the supergravity action
invariant. However, one can view the
resulting theorems as statements about the bosonic spacetimes, and the
spinor field is used as a device.}
\end{footnote}
is given by the standard covariant derivative operator plus terms which
depend on the gauge field strength. Such lowest mass spacetimes are called
Bogomolnyi-Prasad-Sommerfield
(BPS) spacetimes, in analogy with magnetic monopoles that saturate a
similar mass bound \cite{bogo,prasad}
that also follows from a supersymmetric construction \cite{wo}.
These spacetimes have the lowest mass for a fixed value of the charge.
For spacetimes which are asymptotically flat, $|Q|=M$ occurs for extremal
black holes,
or higher dimensional extremal black branes that also have zero temperature.
The non-extremal Riessner-Nordstrom black holes are a nice example of a
known family
of solutions, all more massive than the BPS state $M=Q$.
After ``turning on'' quantum mechanics, one expects that
the higher mass states will evaporate to the ground state by emission
of quantum particles and that therefore BPS spacetimes are quantum
mechanically stable.

Briefly,let's see what this looks like in string theory.
In string theory, two desciptions of states arise.
In perturbative string theory there is a Fock space of states
for a 1+1 dimensional superconformal field theory. This is defined on
the 1+1  dimensional world sheet of the string.
The string propagates in $(9+1)$ dimensional
Minkowski spacetime, or other allowed fixed spacetime geometries. In addition,
perturbative string theory contains Dirichlet-branes, surfaces on which
open strings can
end. D-branes carry a variety of charges.  The states are indexed
by mass, spins, and charges. In another limit of string theory, one
uses a  supergravity field theory description. One thinks of a ``state"
as a spacetime with metric and gauge fields, indexed by mass, angular momenta,
and charges. There is no well defined Hilbert space of quantum states in
this regime.
However, for BPS  perturbative string states, there are many
explicit calculations which display spacetimes that do have the matching
quantum numbers.
D-branes from the perturbative calculations show up in the supergravity
spacetime solutions as black-branes which carry the right kinds of charges.

These BPS spacetimes are the lowest mass states in the
positive mass theorems. In the context of the supergravity end of string
theory, ``excited spacetimes" decay to the lowest mass,
zero temperature configurations by black hole evaporation. This is certainly an
interesting picture. However there are untidy pieces that need explanation.
For example, the family of charged ``black holes" in Anti-deSitter (a
spacetime of constant
negative curvature) are given by (\ref{rnds}) with $\Lambda <0$.
Anti-deSitter spacetimes are
supergravity solutions with maximal supersymmetry.
The lowest mass member, which does have a super-covariantly constant spinor,
is not a black hole but a naked singularity. A naked singularity spells
trouble in classical
general relativity, and it is not
clear how  to think of this object in string theory.

These BPS spacetimes are the lowest mass states in the
positive mass theorems. In the context of the supergravity end of string
theory, ``excited spacetimes" decay to the lowest mass,
zero temperature configurations by black hole evaporation. This is certainly an
interesting picture. However there are untidy pieces that need explanation.
For example, the family of charged ``black holes" in Anti-deSitter (a
spacetime of constant
negative curvature) are given by (\ref{rnds}) with $\Lambda <0$.
Anti-deSitter spacetimes are
supergravity solutions with maximal supersymmetry.
The lowest mass member, which does have a super-covariantly constant spinor,
is not a black hole but a naked singularity. A naked singularity spells
trouble in classical
general relativity, and it is not
clear how  to think of this object in string theory.

Finally, we remind the reader about the puzzle of what are the quantum
mechanical
microstates of black holes that are responsible for their thermodynamic
attributes -
temperature and entropy? A wide range of models have been studied. We will
mention some of
the string theory work. Other appraches include Euclidean quantum gravity
\cite{gh}
 and the entropy of entanglement \cite{bombelli}.
Certain D-brane plus attached open
strings, which are BPS configurations that occur in perturbative string
theory, are
identified with certain types of extremal black holes. The statistical
entropy of
the D-branes plus strings configuration can be computed by explicitly
counting states. In all
cases  where calculations have been done, begining with the work of
\cite{stromvafa}, this has
agreed with the black hole entropy  $A/4$. In the string
picture, Hawking evaporation is modeled as the emission of closed strings
from slightly excited D-branes.  Much work has been
done on computing the low energy excitations of perturbative
D-brane/string BPS configurations. These calculations have been compared
to the presumed corresponding spacetime  black hole evaporation calculations.
Many of the calculations have agreed, and some have
disagreed.
The role of the horizon in defining the black hole entropy is still
a mystery in the string calculations. There is no horizon since these are in
flat spacetime. It is also not clear if the perturbative microstates are in any
sense the same as microstates of the black hole. Nonetheless,
the calculations are very interesting and
understanding black hole thermodynamics continues to be an
area of much current work.

However, when pursuing the definition and attributes of quantum gravity, it is
perhaps well to remember that later understandings often ``...formed just
such  a
contrast with [one's] early opinion on the subject, ...as time is forever
producing
between the plans and decisions of mortals, for their own instruction, and
their neighbor's entertainment." \begin{footnote}{Jane Austen, ``Mansfield
Park".}\end{footnote}

\vskip 0.1in\noindent
{\it Acknowledgments:}
\vskip 0.05in
I would like to thank David Kastor for extensive editing assistance, and Floyd
Williams for organizing this project. This work was supported in part by
NSF grant
PHY98-01875.

\vfill\eject
%%%%%%%%%%%%%%%%%%%%%%

\end{document}